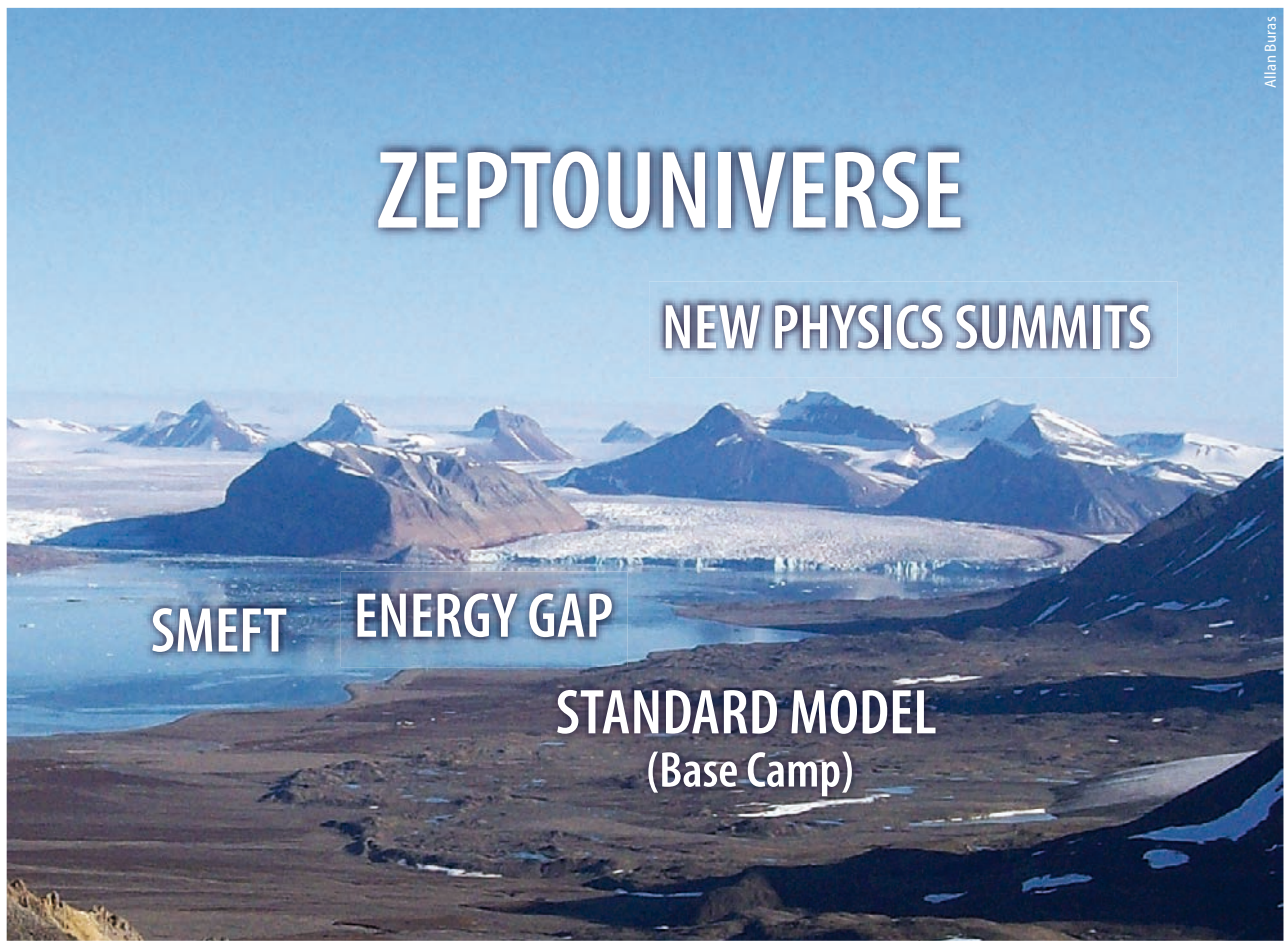

MAX-PLANCK-MEDAILLE

# Expedition to the Zeptouniverse

Flavour experiments promise insights into energy scales as high as 200 TeV and distances as small as $10^{-21}$ meter and offer the chance to identify New Physics.

Andrzej J. Buras

The Large Hadron Collider (LHC) at CERN will directly probe distance scales as short as $10^{-19}$ m, corresponding to energy scales at the level of a few TeV. Presently, higher resolution can only be achieved with the help of quantum fluctuations caused by new particles and new forces that act at very short distance scales and modify the predictions of the Standard Model of particle physics for very rare processes. In this context, weak decays of mesons and leptons play the prominent role besides the transitions between particles and antiparticles in which flavours of quarks and leptons are changed. In this manner, information about the Zeptouniverse corresponding to energy scales as high as 200 TeV or distances as small as $10^{-21}$ m can be obtained.

T he year 1676 was very important for humanity, because Antoni van Leeuwenhoek discovered the empire of bacteria. He called these small creatures *animalcula* (small animals). His discovery was a milestone in our civilization for at least two reasons: He discovered creatures invisible to us which have been killing humans for thousands of years, often responsible for millions of deaths in one year. While Antoni van Leeuwenhoek did not know that bacteria could be dangerous for humans, his followers like Louis Pasteur, Robert Koch and other „microbe hunters" realized the danger coming from these tiny creatures and also developed weapons against this empire [1].

Van Leeuwenhoek was the first human who looked at short distance scales invisible to us and discovered thereby a new underground world. At that time, researchers looked mainly at large distances, discovering new planets and finding laws, such as the Kepler laws which Isaac Newton was able to derive from his mechanics.

While van Leeuwenhoek could reach a resolution of roughly $10^{-6}$ m, this could be improved by twelve orders of magnitude over the last 344 years. On the way down to shortest distance scales, scientists discovered the nanouniverse ($10^{-9}$ m), the femtouniverse ($10^{-15}$ m) relevant for nuclear particle physics and low-energy elementary particle physics and finally the attouniverse ($10^{-18}$ m) which is the territory of contemporary high-energy elementary-particle physics.

Using this overture, I have opened my lecture at the 50th Cracow School of Theoretical Physics held in Zakopane, Poland, in June 2010. At that time, it was strongly believed that the LHC in addition to discovering the only then missing particle of the Standard Model, the Higgs boson, would discover a plethora of new particles, in particular supersymmetric particles or those related to the existence of large extra dimensions like Kaluza-Klein gluons. The Higgs boson was indeed discovered two years later at CERN. But even tremendous efforts of experimentalists and theorists to find New Physics beyond the Standard Model did not result in the discovery of any new particles at the LHC. Thereby, as of August 2020, shifting the masses of supersymmetric particles and Kaluza-Klein gluons significantly above the 1 TeV scale.[1]

Yet, we know that new particles and new forces beyond those present in the Standard Model must exist. The most convincing arguments are based on the following questions, none of which can be answered within the Standard Model:

- What is the dark matter that occupies 27 percent of our universe?
- Why is our universe dominated by matter? This is clearly required for our existence, but the size of the violation of CP symmetry required for the dominance of matter over antimatter soon after the Big Bang is much larger than the one found within the Standard Model.
- Why is the neutron heavier than the proton? This question is significant for our existence.
- What is the origin of neutrino masses and why are they by nine orders of magnitude smaller than the proton mass?
- Why is the mass of the heaviest quark, the top quark, by five orders of magnitude larger than the mass of the lightest quark, the up quark?

It is not the goal of this article to address these questions. Rather, being motivated by them and knowing that New Physics must exist at scales much shorter than explored by now, I would like to concentrate on the following questions:

- Can quantum fluctuations help us with getting some insight into the dynamics at very short distance scales. Could they answer some of these questions, if no direct clear signal of New Physics will be seen at the LHC, i.e., no new particles with masses below 6 TeV will be discovered?
- Can we reach the Zeptouniverse, i.e., a resolution as high as $10^{-21}$ m or energies as large as 200 TeV, by means of quark flavour physics and lepton flavour violating processes in this decade well before this will be possible by means of any collider built in this century?

The photo opening this paper was chosen to illustrate that I am much more optimistic about the future of particle physics than Christoph Wetterich, the winner of Gentner-Kastler-Prize 2019 [2]. Christoph is an esteemed colleague of mine but, in my view, his vision of a desert between the LHC scales and the Planck scale cannot be correct and will be disproved in this decade precisely by flavour physics. In the landscape, photographed by my son Allan during one of his expeditions to the far north, first the Standard Model (our Base Camp) is placed and the energy gap which we are already crossing with the help of the renormalization group equations of the so-called Standard Model Effective Field Theory (SMEFT). However, in order to reach New Physics summits in the far distance, we have to cross the crevasses representing very difficult experiments and difficult theoretical calculations. To this end, we will need brilliant ideas which will guide us through these crevasses so that one day we will reach the summits that will help us to answer at least some of the questions listed above.

After a brief review of the particle content of the Standard Model and of the properties of strong and electroweak interactions described by it, I will present a number of strategies which, with the help of quantum fluctuations, should indeed allow us to get a view of the Zeptouniverse before the advent of future colliders. Subsequently, we will have a brief look at the most interesting anomalies, the departures of the experimental findings from Standard Model expectations. They can be considered as footprints of new particles and new interactions that appear to be beyond the reach of the LHC, although one should not give up the hope that some hints for them will be seen in the next LHC run.

It should be mentioned that this indirect search for new phenomena is by no means new. A classical analogy is the prediction of the existence of the planet Neptune in 1846 by Urbain Le Verrier based on an anomaly in the orbit of the Uranus. Neptune's existence was soon after confirmed by the German astronomer Johann Gottfried Galle. The gauge bosons $W^\pm$ and $Z^0$ were seen indirectly in lower-energy experiments well before their discovery in 1983 at CERN. The presence of the $W^\pm$ bosons was actually felt already in the 1930s in the context of the Fermi theory of $\beta$-decays even if only in the 1960s theorists realized what are the elementary particles mediating these decays. $Z$ boson was seen first in 1973 through the discovery of neutral currents at CERN – ten years before this boson was discovered also at CERN.

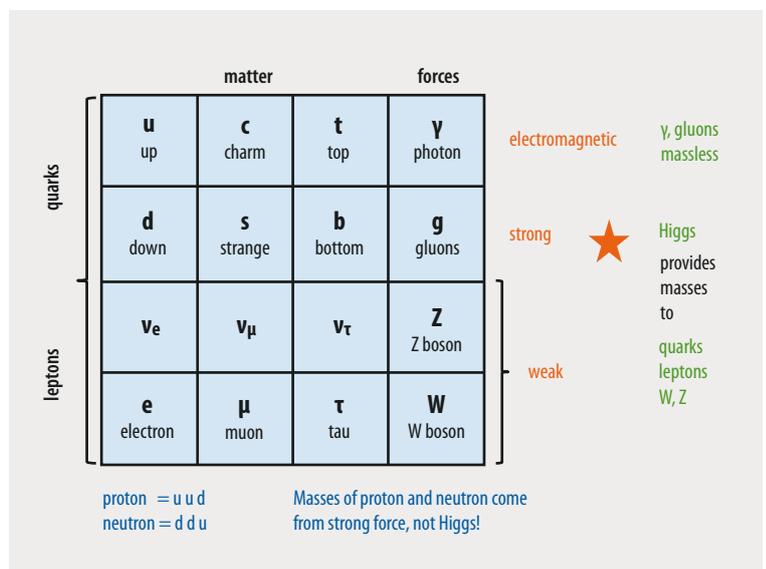

**Fig. 1** The particle content of the Standard Model

---

1) Of course, one cannot exclude the existence of very light particles, like axions, that being very weakly coupled to standard matter could not be detected until now.

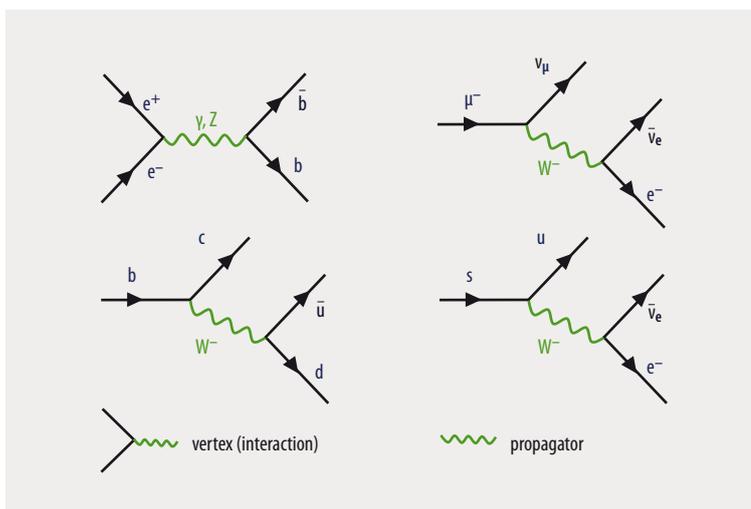

**Fig. 2** The world of Feynman Diagrams: more complicated diagrams can be constructed and probabilities for a given process evaluated.

## Express Review of the Standard Model

In the Standard Model matter consists of four fermion families (**Fig. 1**): up quark ($u,c,t$), down quark ($d,s,b$), neutrino ($\nu_e, \nu_\mu, \nu_\tau$) and electron ($e, \mu, \tau$). They correspond to the four rows in the table. Particles in a given family have the same quantum numbers, in particular similar electric charges: 2/3, –1/3, 0 and –1, respectively. Their antiparticles have opposite electric charges. The members of a given family can only be distinguished by their masses which increase in the table from left to right. The columns in this table correspond to three generations of these elementary particles and the different names of quarks and leptons are called flavours: six quark flavours and six lepton flavours.

The interactions between the matter fields are mediated by gauge bosons with spin 1. For strong interactions (Quantum Chromodynamics, QCD), these are the electrically neutral massless gluons ($G^a$). For the electroweak interactions (Quantum Flavourdynamics) these are the electrically neutral massless photon ($\gamma$), the electrically neutral heavy $Z$ boson and finally the charged heavy $W^\pm$. The masses of all these particles, according to the Standard Model, are generated through their interactions with an electrically neutral scalar particle (Spin 0), the Higgs boson.

The strong and electroweak interactions have a number of properties encoded in the Lagrangian of the Standard Model that can be found in any textbook on particle physics. The following four properties of these interactions (**Fig. 2**) will be relevant for us:

■ By themselves, neutral gauge bosons ($G^a, \gamma, Z$) and the Higgs boson cannot transfer a fermion with one flavour into another fermion with a different flavour. There are simply no vertices in any Feynman diagram that involve any of these bosons and two fermions (quarks or leptons) with different flavours. For instance, a vertex with $Z$ boson, $b$ quark and a $s$ quark does not exist. This is assured by the so-called Glashow-Iliopoulos-Maiani (GIM) mechanism. Needless to say, Standard Model gauge bosons cannot transfer a lepton into a quark or vice versa. This is already evident from charge conservation. The interactions mediated by these bosons also conserve parity (P), charge conjugation (C) and CP-parity. The charged gauge bosons $W^\pm$ change flavour and violate maximally parity and charge conjugation implying that only left-handed quarks and left-handed leptons take part in the charged current weak interactions. These interactions are parametrized in the case of quarks by the unitary Cabibbo-Kobayashi-Maskawa (CKM) matrix which depends on three real parameters (the so-called mixing angles) and one complex phase responsible for CP-violation in the Standard Model. In the case of leptons, $W^\pm$ interactions are parametrized by the unitary Pontecorvo-Maki-Nakagawa-Sakata (PMNS) matrix which can have two additional phases relative to the CKM matrix, the so-called Majorana phases, related to special properties of neutrinos, the only neutral fermions we know.

■ The gauge interactions mediated by neutral $G^a$, $Z$ and $\gamma$ are universal in a given family. In particular the interactions of $Z$ with $e$, $\mu$ and $\tau$ are the same.

■ While electroweak interactions are weak and can be calculated within perturbation theory, the strong interactions are strong at scales below 1 GeV in order to bind quarks inside hadrons like mesons, the proton and the neutron. At these scales, only non-perturbative methods are useful. These are in particular the numerical Lattice QCD and analytical methods like Dual QCD and Chiral Perturbation Theory. For scales above 1 GeV, strong interactions are sufficiently weak due to the property of asymptotic freedom in QCD, so that their effect can be calculated within perturbation theory. Yet, in the presence of vastly different energy scales, like the hadronic scale $\mathcal{O}(1\,\text{GeV})$, the electroweak scale (246 GeV) and New Physics scales often well above 1 TeV, the appearance of large logarithms of the ratios of these scales multiplying the gauge couplings requires their summation to all order of perturbation theory. To this end, very efficient renormalization group methods, known also in the field of phase transitions, are used. They are discussed in much more detail in my recent book [3].

We have seen that, within the Standard Model, neutral gauge bosons are not able to change flavour by themselves. However, one can construct complicated Feynman diagrams (loop diagrams) involving these bosons together with $W^\pm$ that do change flavour. One example of such a diagram is called penguin diagram (**Fig. 3**). But such loop diagrams can have a different shape like the so-called box diagrams, in which the neutral gauge boson is replaced by the pair $W^+$ and $W^-$ that has zero charge. Processes of this type are called Flavour Changing Neutral Current (FCNC) processes and are very important for our discussion. Prominent examples are the decays $B_s^0 \to \mu^+\mu^-$, $K^+ \to \pi^+\nu\bar\nu$ and particle-antiparticle transitions like ($B_{s,d}^0 - \bar B_{s,d}^0$) mixings.

According to the usual Feynman-diagram calculus, the probability for a given process to occur is proportional to the product of couplings present in the vertices of the diagrams governing this process. Involving several weak couplings and often small elements of the CKM matrix present in the vertices in one-loop diagrams, FCNC processes are strongly suppressed within the Standard Model. This suppression is partly lifted if the top-quark is present inside the loop and the contributions of such diagrams grow quadratically with the top-quark mass. This is only the case with the decays of $K$ and $B$ mesons but not for the $D$ mesons and leptons, so that most interesting measurements of FCNCs to date come from weak decays of $K$ and $B$ mesons, ($B_{s,d}^0 - \bar B_{s,d}^0$) mixings and ($K^0 - \bar K^0$) mixing. But as these processes are very strongly suppressed in the Standard Model, they are more powerful in the search for New Physics than processes that are possible within the Standard Model through the exchange of a simple $W^\pm$ (**Fig. 2**). In this way, the Standard Model contributions represent a significant background in the search for New Physics.

On the other hand, and beyond the Standard Model, the GIM mechanism is often absent and FCNC processes can be governed by simpler diagrams involving new heavy particles. For example, a very heavy neutral gauge boson $Z'$ contributes to ($B_{s,d}^0 - \bar B_{s,d}^0$) mixing (**Fig. 4**). Although its propagator ($1/M_{Z'}^2$) strongly suppresses this contribution, the absence of several weak couplings relatively to one-loop diagrams partly lifts this suppression. Consequently, such contributions can be relevant and play often significant roles in finding new phenomena as we will explain now. Quantum fluctuations involving new particles can also generate flavour changing vertices in which $Z$ bosons changes an $s$ or $d$ quark into the $b$ quark (**Fig. 4**).

### The Technology to Reach the Zeptouniverse

Main players in indirect searches for New Physics are presently the mesons, quark-antiquark bound states

$B_d^0 = (\bar b d)$, $B_s^0 = (\bar b s)$, $B^+ = (\bar b u)$, $K^+ = (\bar s u)$, $K_L = (\bar s d)$, $\pi^+ = (\bar d u)$,

but also leptons and mesons with charm quark will surely play important roles in the search for New Physics in this decade. In order to reach high resolution at short distance scales, one has to produce many of these mesons in high energy collisions. They subsequently decay into lighter particles. The goal of experimentalists is then to measure very accurately the probability with which a given meson decays into a particular final state such as $\mu^+\mu^-$, $\pi^+\nu\bar\nu$, $\pi^+\pi^-$ among many possible final states. These probabilities, normalized to unity (100 percent), are called branching ratios.

The goal of theorists is to calculate these with high precision within the Standard Model and compare them with the experimentally measured branching ratios. Any difference between the experimental branching ratio and the one predicted by the Standard Model is a hint at the existence of new particles that are often too massive to be produced at the LHC. But through quantum fluctuations represented by propagators in Feynman diagrams they can affect various branching ratios so that the latter can differ from the ones predicted in the Standard Model.

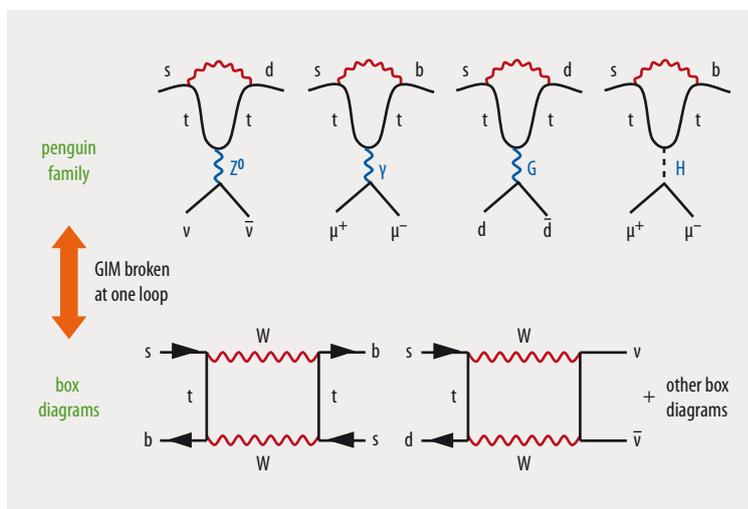

**Fig. 3** Penguin diagrams and box diagrams

To identify new particles in this indirect manner, it is crucial to test many different branching ratios for the meson decays listed above. The Particle Data Group (PDG) collects the experimental values of these branching ratios and this collection amounts to thousands of different numbers. In the case of lepton decays, only upper limits on the branching ratios are known, because to date, no FCNC process in the lepton sector has been observed experimentally. Yet, they must exist at a certain level because of non-vanishing neutrino masses. However, these masses are tiny, and in the Standard Model, such processes are predicted to have very small branching ratios like $\mu \to e\gamma$ in the ballpark of $10^{-54}$. Any observation of such decays would be a clear signal of New Physics.

As far as flavour expedition to the Zeptouniverse is concerned, only a fraction of the branching ratios collected by the PDG is of interest to us. These are the ones which are predicted to be very small in the Standard Model, because then the background from the Standard Model does not prevent seeing new phenomena. Very small does not mean 0.01 or 0.001 but often $10^{-6}$, $10^{-9}$ and even $10^{-11}$ in the case of meson decays. For leptons, such branching ratios are even smaller as already stated above. For instance, the present experimental upper bound on the branching ratio of $\mu \to e\gamma$ amounts to $10^{-13}$ and it is amazing that experimentalists can measure such tiny values.

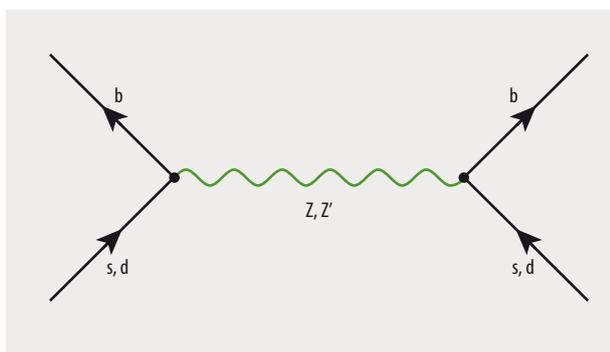

**Fig. 4** This Feynman diagram shows a FCNC process mediated by a heavy neutral gauge boson $Z'$ and/or $Z$.

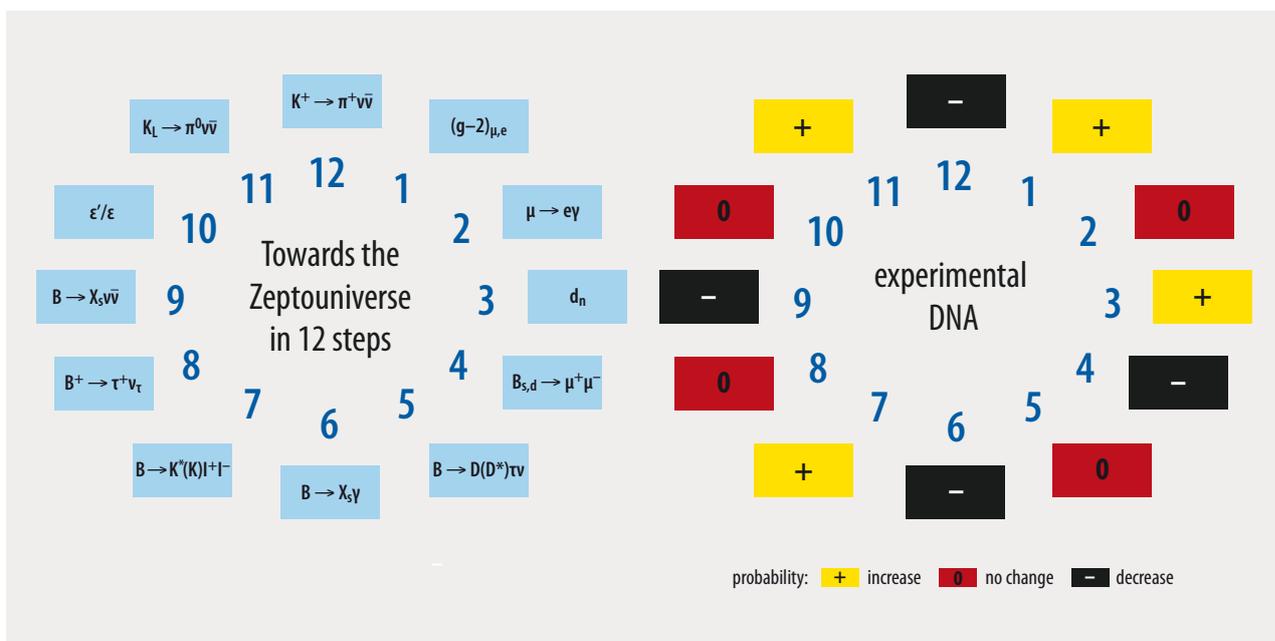

**Fig. 5** General Structure of DNA: considered processes (left) and experimental DNA (right).

Let us assume that one day we will have hundreds of very precise measurements of various branching ratios for decays of mesons and leptons and very precise Standard Model predictions for them. This will allow us to construct a series of differences between experimental and theoretical branching ratios calculated in the Standard Model. These differences will be generally positive, negative or consistent with zero. A positive difference means that there is a New Physics contribution enhancing the branching ratio, while a negative one signals New Physics which suppresses the branching ratio relative to the one predicted by the Standard Model. An example is given in **Fig. 5** where a selected number of processes is shown compared to the coding for differences between Standard Model predictions and experimental data found in the possible future.

These twelve examples are representatives of the hundreds of branching ratios at our disposal one day, among them many that will correspond to yellow or black colours thereby indicating the presence of New Physics. This is similar to having a DNA of a criminal, represented here by New Physics, and our goal is to identify him or her in the most efficient manner. Of course, there are many criminals behind these numbers, in general complicating this search.

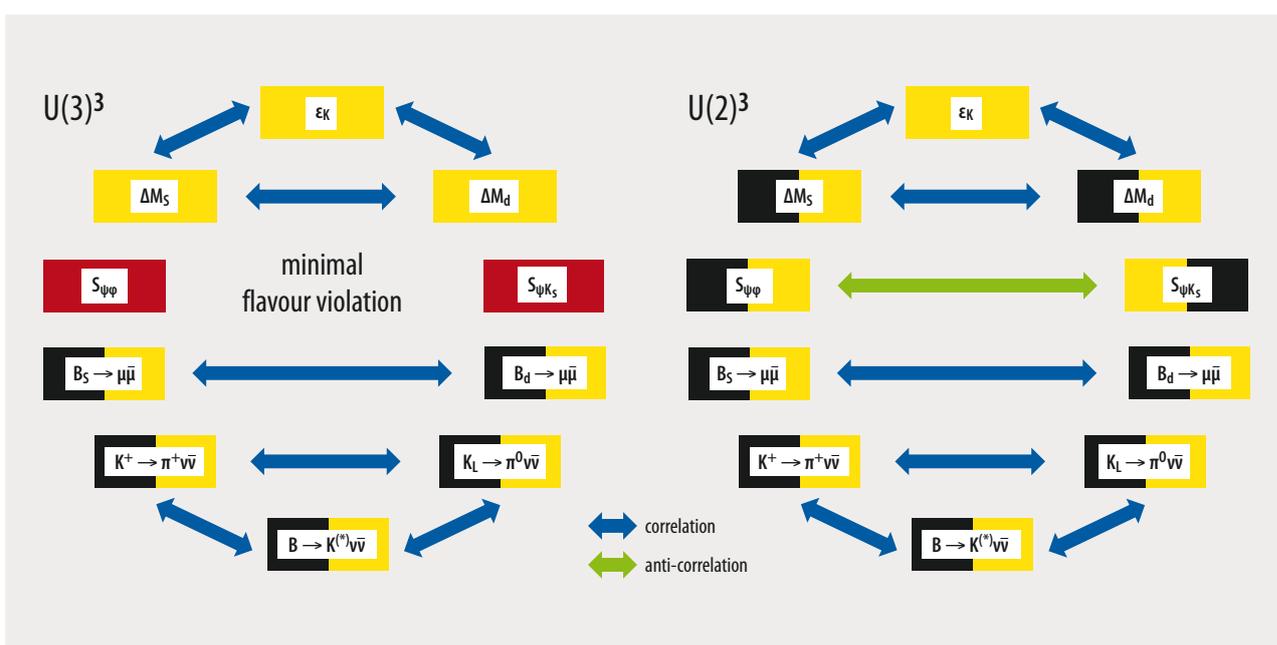

**Fig. 6** DNA-chart of Minimal Flavour Violation models and DNA-chart of $U(2)^3$ models

Without any theoretical input, this goal cannot be realized. In addition to finding these new particles, we would obviously like to know their interactions which could again be mediated by gluons, photons and a *Z* boson. But generally new Feynman diagrams will include additional gauge bosons, new fermions and scalars that is brothers and sisters of Standard Model particles which could be both electrically neutral or charged. One restriction comes from the observation that for energy scales much larger than the scale of spontaneous breakdown of the Standard Model gauge symmetry $SU(3)_C \times SU(2)_L \times U(1)_Y$, this symmetry must be exact which reduces the number of possible theories. Yet, without additional dynamical assumptions, the New Physics contributions in full generality turn out to depend on 1350 real parameters and 1149 complex phases. This number becomes smaller if one considers a specific class of processes, but still it is not possible to determine all these parameters in low-energy experiments without some specific simplifying assumptions about the structure of New Physics. This fact is fortunately key to reduce the number of possibilities for physics beyond the Standard Model.

In my view the most efficient strategy in this context is to consider many concrete New Physics models or some simplified versions of them and calculate within each of them as many observables as possible. Most of my younger colleagues would then just put these results into a computer code to calculate $\chi^2$ or a *p* number for each model. Others would present multi-dimensional plots in the space of the parameters of the model to identify the range of parameters that is ruled out by experiment, often leaving small oases where a given model can still survive. These are valid procedures, but in my view not sufficiently transparent as far as the nature of the New Physics we are searching for is concerned. In my view, it is better to first construct for each model its DNA consisting of +, 0 and – as explained above, and then compare it to the DNA of the criminal(s) determined by experiment as given in **Fig. 5**.

However, it is rare for a given observable in a given theory to be uniquely suppressed or enhanced relative to the Standard Model. Frequently, two observables are correlated or uncorrelated with each other. Thus, the enhancement of one observable implies uniquely an enhancement (correlation) or suppression (anticorrelation) of another observable. Among further possibilities, it can also happen that a change in the value of a given observable implies no change in another observable. After adjusting the parameters of a given theory in order to reproduce the enhancement of a given branching ratio, this theory predicts also enhanced or suppressed values for other observables and sometimes there is no effect of New Physics on some other branching ratios.

Applying then the information from a given theory requires sometimes further significant theoretical work. The strategy, developed in collaboration with Jennifer Girrbach-Noe [4], is to connect a given pair of branching ratios that are correlated or anticorrelated with each other by a line in a DNA-chart. The absence of a line means that two given observables are uncorrelated.

This strategy is illustrated with four simplified models discussed in detail in [3, 4]. In the left part of **Fig. 6**, we show the DNA-chart of the so-called Minimal Flavour Violation New Physics scenario which is based on the flavour symmetry $U(3)^3$. It is the minimal extension of the Standard Model. The right part of this figure shows the DNA of models with reduced flavour symmetry $U(2)^3$. **Fig. 7** shows two theories with a heavy neutral *Z′*, which only interacts with left-handed and right-handed quarks, respectively. This two charts can also represent the Standard Model $Z$ boson, which acquired flavour-changing interactions through New Physics. It should be noted that all four charts differ from each other. As soon as experiments favour one model, one can begin to investigate it in more detail within the chosen theory, calculating $\chi^2$ and $p$-numbers. If neither is favoured by nature, new DNAs for other theories have to be constructed.

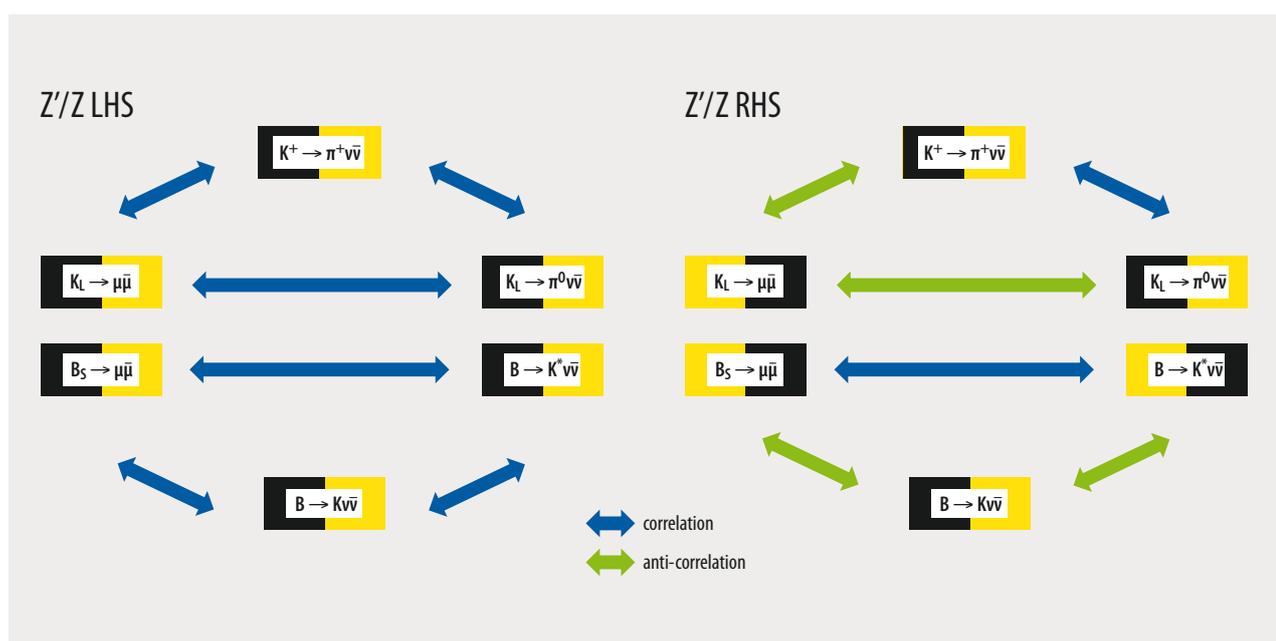

**Fig. 7** DNA-charts of *Z′* models with LH and RH currents

### Anomalies

During the last ten years a number of deviations from Standard Model predictions have been measured by experimentalists. The most popular ones are presently the so-called $B$-physics anomalies observed in the BaBar, LHCb, CMS, ATLAS, and Belle experiments. The observed data imply that the lepton flavour universality breaks down, i.e., the branching ratios for rare decays of $B$ mesons with muons and electrons in the final state differ from each other by 2 to 3 standard deviations. A similar phenomenon is observed when a $B$ meson decays into muons compared with decays into $\tau$ leptons. The prime criminals behind such phenomena are leptoquarks, heavy bosons with spin 0 or 1 changing quarks into leptons or vice versa. In the case of the violation of the $\mu - e$ universality, a heavy gauge boson $Z'$ could be responsible for this anomaly as well. Yet, we still have to wait for more precise data and in some cases for more precise calculations to be confident that these effects are indeed more than statistical fluctuations.

Another anomaly found already twenty years ago at the Brookhaven National Laboratory is the deviation of the anomalous magnetic moment of the muon, $(g-2)_\mu$, from its rather precise Standard Model value. A new experiment at FNAL and an independent experiment at J-PARC in Japan should clarify whether new physics is hidden in this finding. Again, leptoquarks could be responsible for this anomaly.

The study of the violation of CP symmetry in $K \to \pi\pi$ decays represented by the so-called ratio $\epsilon'/\epsilon$ is also of considerable importance. Unfortunately, the calculation of this ratio is subject to large non-perturbative uncertainties. Presently, we do not know whether new phases beyond the CKM phase are necessary to explain the existing data from NA48 at CERN and KTeV at Fermilab, known to us already for twenty years. Such new phases, if required, could in principle explain the dominance of the matter over the antimatter in the universe. If that is the case the criminals among other possibilities could be a heavy $Z'$ and heavy vector-like quarks in which left and right components transform identically under $SU(2)_L$. Leptoquarks turn out to be less useful in this case: Even the explanation of a moderate $\epsilon'/\epsilon$ anomaly would imply very large branching ratios for rare Kaon decays, which are experimentally excluded. This result highlights the importance of correlations between various processes. Such correlations should be very powerful when the rare decays $K^+ \to \pi^+ \nu\bar\nu$ and $K_L \to \pi^0 \nu\bar\nu$ will be accurately measured by the NA62 experiment at CERN and by the KOTO experiment in Japan, respectively. In addition, the CLEVER experiment at CERN should contribute in the future to the study of $K_L \to \pi^0 \nu\bar\nu$.

### Conclusions

Detailed studies of the ability of flavour physics to provide information about new physics beyond 100 TeV have been performed in [5, 6]. They show that the particle-antiparticle mixings are most efficient to reach these scales [5], but a detailed picture of the Zeptouniverse can only be obtained through the study of rare $K$ and $B_{s,d}$ decays [6] and lepton flavour violating decays like $\mu \to e\gamma$ and $\mu^- \to e^- e^+ e^-$ [7]. Additionally, electric dipole moments of the neutron and of various atoms are very important in the search for new phenomena [8].

There is no question that flavour physics has a great future through experiments in Europe, Japan and USA. I am looking forward to the year 2030, when hopefully a concrete picture of the Zeptouniverse will be known.

## The Author

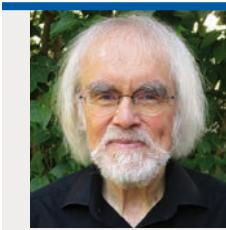

**Andrzej J. Buras** studied physics in Warsaw. He received his doctorate in 1972 at the Niels Bohr Institute (Copenhagen). Postdoctoral years at CERN (Geneva), Fermilab (Chicago) and SLAC (Stanford University) were followed by a position at the Max Planck Institute for Physics in Munich. In 1988, he was appointed full professor at the Physics Department of the TU Munich. Since 2012, he is Emeritus of Excellence of the TUM and heads a group on Fundamental Physics at the Institute for Advanced Study of the university. Andrzej J. Buras was awarded the Smoluchowski-Warburg Medal of the German and Polish Physical Societies in 2007. From 2011 to 2016, he was funded by an ERC Advanced Grant. He is a full member of the Bavarian Academy of Sciences (2010), foreign member of the Polish Academy of Sciences (2013) and of the Polska Akademia Umiejętności in Cracow.

**Prof. Dr. Andrzej J. Buras**, TUM Institute for Advanced Study, Lichtenbergstraße 2a, 85748 Garching, Germany